\documentclass[aps,prl,twocolumn,amsmath,amssymb,floatfix,superscriptaddress,showpacs]{revtex4}
\usepackage{graphicx}

\begin{document}


\title{Voltage induced control and magnetoresistance of noncollinear frustrated magnets}


\author{A.~Kalitsov}
\affiliation{SPINTEC, UMR 8191 CEA/CNRS/UJF, 38054 Grenoble, France}
\author{M.~Chshiev}
\affiliation{SPINTEC, UMR 8191 CEA/CNRS/UJF, 38054 Grenoble, France}
\author{B.~Canals}
\affiliation{Institut N\'eel, CNRS-UJF, BP 166, 38042 Grenoble Cedex 9, France}
\author{C.~Lacroix}
\affiliation{Institut N\'eel, CNRS-UJF, BP 166, 38042 Grenoble Cedex 9, France}


\date{\today}

\begin{abstract}
Noncollinear frustrated magnets are proposed as a new class of spintronic materials with high magnetoresistance which can be controlled with relatively small applied voltages. It is demonstrated that their magnetic configuration strongly depends on position of the Fermi energy and applied voltage. The voltage induced control of noncollinear frustrated materials (VCFM) can be seen as a way to intrinsic control of colossal magnetoresistance (CMR) and is the bulk material counterpart of spin transfer torque concept used to control giant magnetoresistance in layered spin-valve structures.
\end{abstract}

\pacs{72.15.Eb, 72.15.Gd, 75.10.Lp, 75.10.Jm, 75.47.-m, 75.47.Pq}
\maketitle

The discovery of giant magnetoresistance (GMR)~\cite{Baibich,Binash} in magnetic multilayered structures has generated
a new field of spin-based electronics~\cite{Fert,Wolf}, or spintronics, which combines two traditional fields of physics: 
magnetism and electronics. A spin-valve concept~\cite{Dieny} used in GMR structures allows controlling the magnetic 
configuration of its ferromagnetic layers by (i) application of relatively small magnetic fields or 
(ii) passing spin polarized currents using spin transfer torque (STT)~\cite{Slonczewski,Berger}. These factors made them ideal 
systems for spintronic applications such as magnetic random access memories (MRAM) and magnetic field sensors used in read heads. 

The advent of GMR has considerably increased an interest in related phenomenon in bulk materials, 
colossal magnetoresistance (CMR)~\cite{Helmolt,Jin}, which is several orders higher than GMR and unlike the latter, 
can be viewed as an "intrinsic" property of material itself. To date, the CMR is typically observed in certain manganite compounds 
with the bulk magnetic configuration controlled by applying the magnetic field (similar to method (i) mentioned above for spin valves) 
but requires characteristic magnetic fields of several Tesla~\cite{Jin}. Such high fields make them inappropriate for use in 
spintronic applications where appropriate scale should be about Oersteds. However, one may expect the possibility of controlling 
the intrinsic magnetic configuration of the bulk materials (and thus of CMR) by passing spin polarized currents through them similar to 
STT mechanism (ii) mentioned above for spin-valves. Since the STT in the latter originates from non-collinearity of their adjacent
magnetizations, the same requirement should hold for the bulk materials. 

Here we promote for the first time magnetically frustrated bulk materials as a new paradigm for spintronic applications 
with high magnetoresistance which can be controlled with relatively small applied voltages and does not require spin polarized currents. 
This novel phenomenon may be viewed as the "bulk" counterpart of STT in layered spintronic structures (spin valves) and may represent a crucial interest 
both from applications and fundamental points of view. Below we demonstrate that the magnetic configuration 
of the bulk frustrated material is changed under applied voltage leading to the strong variation of its conductance. 
\begin{figure}
\includegraphics[width=7cm]{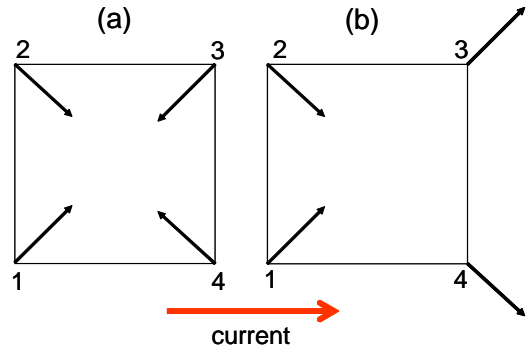}
\caption{\label{fig1} (a) "4in" and (b) "2in2out" configurations for square lattice}
\end{figure}

 
The key mechanism at stake is a somewhat microscopic equivalent of spin torque, allowing for local 
spin flips within the microscopic spin texture. Those local moves requires non collinearity of the spin configuration, which would otherwise 
be insensitive to the current. Furthermore, in order to avoid a continuous response of the spin configuration, we require the 
ground states to be well defined minima, such that the switching from one to the other looks like 
a logical operation. These two aspects are well taken into account as soon as the underlying localized moments 
possess strong multiaxial anisotropies, in order to behave bitwise like and fulfill the non 
collinearity criterion. Several Rare-Earth-transition metal intermetallics systems are relevant candidates : 
in these systems frustration arises due to the competition of crystal field anisotropy, 
exchange and quadrupolar interactions (for a review see Ref.~\cite{gignoux1991}). 
These intermetallics systems often show a non-collinear magnetic structure: 
this is the case for example of TbGa$_2$\cite{auneau1995}, HoGe$_3$\cite{schobinger2008} 
or Uranium compounds\cite{burlet1981}. Naturally, pyrochlores in which frustration is due to the crystal 
structure are also relevant candidates, as it is well known that in such exotic systems, 
non collinear low temperature magnetic phases are often stabilized\cite{gardner2009}, 
among those, few metallic compounds have been identified like Pr$_2$Ir$_2$O$_7$\cite{nakatsuji2006,Machida2007} 
and the family of the pyrochlore Molybdates,  R$_2$Mo$_2$O$_7$\cite{kezsmarki2004}. 
\begin{figure}
\includegraphics[width=7cm]{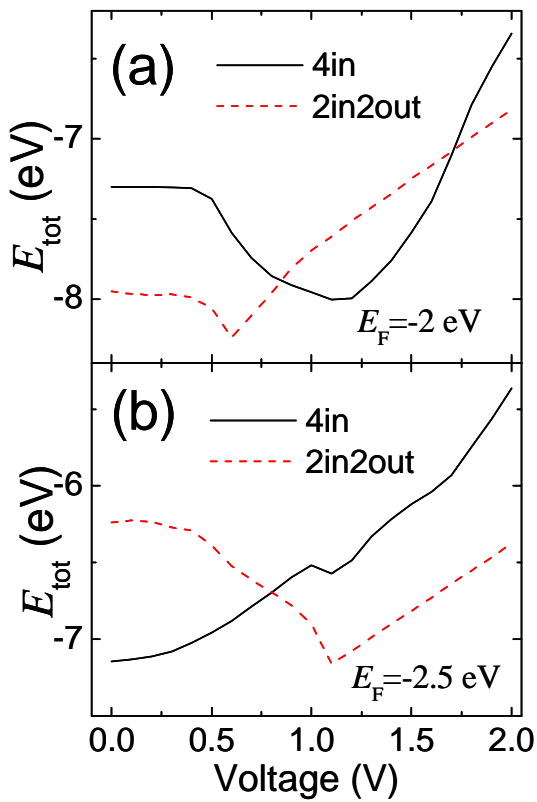}
\caption{\label{figVOLTAGE} (a) and (b): total energy as a function of applied voltage for $E_F$=-2~eV and  
$E_F$=-2.5~eV, respectively, shown by arrows in Fig.~\ref{figDOS}.}
\end{figure}

For argument's sake, we have used a two-dimensional square lattice model in order to focus on the driving mechanism that is at stake and to describe 
its qualitative behavior. Realistic realizations would involve more complicated structures, like the one in the 
3-D pyrochlores~\cite{gardner2009}. We consider a square lattice of classical localized moments ${\bf S}_i$ with strong local 
on-site uniaxial anisotropy ($D_0$) along square diagonals (represented by unit vector ${\bf n}_i$) and coupled 
through intersite exchange $I_{ij}$. In addition, moments ${\bf S}_i$ are coupled with conduction electrons through the 
local exchange ($J_0$) described here quantum mechanically using tight-binding model. Unlike collinear systems, 
non-collinearity provided by magnetic frustration is a key ingredient for switching 
phenomenon proposed here since its origin is due to STT mechanism acting locally by conduction electrons on 
localized moments ${\bf S}_i$. The Hamiltonian of the system has the form:
\begin{eqnarray}
\hat{H} = -\sum\limits_{i,j} I_{ij}\mbox{\boldmath$S$}_i \cdot \mbox{\boldmath$S$}_j
- D_0 \sum\limits_{i} \left( \mbox{\boldmath$n$}_i \cdot \mbox{\boldmath$S$}_i \right)^2 + \nonumber \\
t \sum\limits_{i,j,\sigma} \left( \hat{c}^{\dagger \sigma}_i \hat{c}^{\sigma}_j + H.c. \right) 
- J_0 \sum\limits_{i} \hat{c}^{\dagger \alpha}_i \left( \mbox{\boldmath$\sigma$}_{\alpha \beta} 
\cdot \mbox{\boldmath$S$}_i \right) \hat{c}^{\beta}_i,
\end{eqnarray}
where $I_{ij}$ and $J_0$ are the intersite and the local exchange constants respectively, $D_0$ is the
uniaxial anisotropy constant, t is the hopping integral between two neighboring sites, 
$\hat{c}^{\dagger \sigma}_i$ and $\hat{c}^{\sigma}_i$ are the creation and annihilation operators
of the conduction electron with the spin $\sigma$ on site $i$ and ${\boldmath \sigma}_{\alpha \beta}$
is the vector of Pauli matrices. For the chosen model, nearest neighbor interactions $I_1$ are irrelevant (because nearest neighbor spins are always orthogonal)
so that only the second nearest neighbor one $I_2$ is taken into account. We emphasize that dealing with longer ranges interaction does not affect 
further results reported here since only energy differences between magnetic configurations matter.

In the limit of low temperature and $D_0 \rightarrow \infty$ the localized moments ${\bf S}_i$ are 
strictly collinear with ${\bf n}_i$ (${\bf S}_i \| {\bf n}_i$) 
and the total energy of the system within a constant shift becomes
\begin{eqnarray}\label{Etot}
E_{tot} = -I_2 \sum\limits_{i,j} \mbox{\boldmath$S$}_i \cdot \mbox{\boldmath$S$}_j 
+ Tr[\hat{H} \hat{\rho}] = \nonumber \\ 
= -I_2 \sum\limits_{i,j} \mbox{\boldmath$S$}_i \cdot \mbox{\boldmath$S$}_j
- \frac{i}{2\pi} \sum\limits_j \int E G_{jj}^< (E) dE,
\end{eqnarray}
where $G^<$ and $\hat{\rho}$ represent the "lesser" non-equilibrium Green function and density matrix respectively~\cite{Lifshitz}, 
and the Green function indices include spin index as well.

We adopt a conventional transport approach where the system is considered to consist of three regions: semi-infinite left ($L$) and right ($R$) regions
connected to the middle scattering region ($M$) which is assumed to be very long. Despite of all three regions being identical, such a subdivision is chosen in order to consider all non-equilibrium processes occuring in the scattering region $M$ while the left and right regions are in thermodynamical equilibrium described with Fermi-Dirac distribution functions  $f_{L(R)} = f(E-\mu_{L(R)})$, where $\mu_{L(R)}$ is the chemical potential in the left(right) region. The "lesser" Green function can be written as     
\begin{equation}\label{Gless}
G_{ij}^< = i (f_L \psi_i^L\psi_j^{L*} + f_R \psi_i^R\psi_j^{R*}),
\end{equation}
where $\psi_j^{L(R)}$ is the single-electron wave function on site $j\in M$ incident from the left (right). It is straightforward from the definition 
of retarded and advanced Green functions that
\begin{equation}\label{GaGr}
G_{jj}^a - G_{jj}^r = i (\psi_j^L\psi_j^{L*} + \psi_j^R\psi_j^{R*}).
\end{equation}
Since the middle region is assumed to be very long, the finite potential drop $eV=\mu_{L}-\mu_{R}$ 
results into infinitesmal change from one unit cell to another within the middle region $M$.
In this case the reflections from the $L|M$ and $R|M$ boundaries are negligible and one can write for 
the left(right) wave functions $\psi_j^{L(R)} = c_{L(R)} \exp{(+(-) k_x a j)}$, where $k_x$ is the wave vector along $x$-axis and $a$ is the lattice constant. 
One can show that the expression for the charge current which is proportional to $(f_L-f_R)$~\cite{Meir} is satisfied when $|\psi_j^L|^2 = |\psi_j^R|^2$.
Indeed, the charge current can be expressed using the lesser Green function as~\cite{Caroli}:
$$
I \sim \int {[(G_{j+1,j}^{<}  - G_{j,j+1}^{<})]dE}.
$$
Using Eq.~(\ref{Gless}) and aforementioned wave functions it yields the following form:
$$
I \sim \int {\sin{k_x a}\left[f_L |c_L|^2-f_R |c_R|^2 \right]dE}
$$
which is proportional to $(f_L-f_R)$ only when $|c_L|^2=|c_R|^2$ (and consequently $|\psi_j^L|^2 = |\psi_j^R|^2$).

Then, it straightforwardly follows from Eq.~(\ref{Gless}) and Eq.~(\ref{GaGr}) that 
\begin{equation}
G_{jj}^<(E) = i(f_L + f_R) \Im(G_{jj}^r(E))
\end{equation}
where $\Im$ notation for imaginary part is used.
The total energy given by expression~(\ref{Etot}) of the system per a unit cell yields:
\begin{eqnarray}\label{finalE}
E_{tot} = -I_2 \sum\limits_{i,j} \mbox{\boldmath$S$}_i \cdot \mbox{\boldmath$S$}_j - \nonumber \\
\frac{1}{2\pi} \int \left(f_L + f_R \right) \sum\limits_{j} \Im(G_{jj}^r(E)) E dE,
\end{eqnarray}
where $G_{jj}^r(E)$ is calculated as $j$th diagonal element of the matrix $[E-\hat{H}+i\delta]^{-1}$.
Thus, the magnetic configuration corresponding to the minimum of the total energy is resulting from the interplay 
of intersite exchange interaction between localized moments on the next nearest neighbour sites (the first term in Eq.~(\ref{finalE}))
and the exchange interaction between the localized moment and the conduction electron described by the second term in Eq.~(\ref{finalE}). 

For simplicity, among the possible magnetic configurations of the system (${\bf S}_i \| {\bf n}_i$) we select "4in" and "2in2out" ones which are the two configurations with the largest energy separation in this example~(see Fig.~\ref{fig1}). 
We choose for parameters $J_0 = 2~eV$, $t=1~eV$ and $I_2 = 0.1t$~\cite{Solovyev}. 

In Fig.~\ref{figVOLTAGE} we show the dependence of the total energy for the two aforementioned selected configurations as a function of applied voltage. 
First, one can note that $E_{tot}$ strongly depends on the Fermi level position which is used to represent different materials. Indeed, 
when the system is in equilibrium, its "2in2out" ("4in") state is energetically more favorable for $E_F=-2~eV$ ($-2.5~eV$) as shown in Fig.~\ref{figVOLTAGE}(a) and (b), respectively. Furthemore, the strong variation of the total energy as a function of applied voltage causes the system to switch from an equilibrium state to another one at certain critical voltage (see Fig.~\ref{figVOLTAGE}). Moreover, further increase of the applied voltage may again reverse the state of the system (Fig.~\ref{figVOLTAGE}a).
This demonstrates that the magnetic configuration of the system can be controlled by the applied voltage. 

\begin{figure}
\includegraphics[width=7cm]{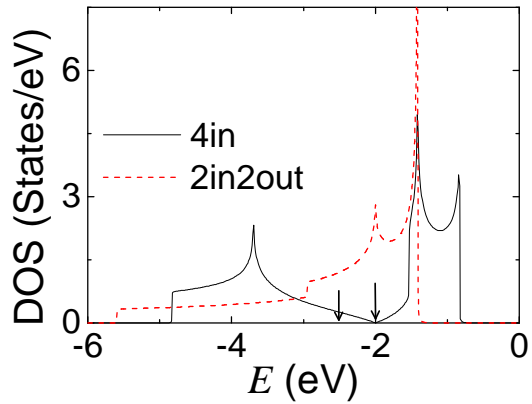}
\caption{\label{figDOS} Total density of states for "4in" (black solid) and "2in2out" (red dashed) configurations 
represented in~Fig.~\ref{fig1}.}
\end{figure}

The mechanism of these dependences may be understood from the corresponding total density of states (DOS)~\cite{DOSdefin} for "2in2out" and "4in" configurations represented 
in Fig.~\ref{figDOS} where for convenience purposes only negative energy range is shown since DOS($E$)=DOS($-E$). The DOS for both magnetic states 
have sharp peaks and band gaps causing strong dependence of the preferable configuration on the Fermi level $E_F$ both in and out of equilibrium. 
Indeed, as on can see from the second term in Eq.~(\ref{finalE}), 
the total energy is defined by the sum of two products of the DOS with $E$ integrated till $\mu_L$ and $\mu_R$. 
In the absence of applied voltage, $E_{tot}$ is calculated with $\mu_L=\mu_R=E_F$, i.e. defined by the position of the Fermi level indicated 
by black arrows in Fig.~\ref{figDOS} corresponding to two values $-2~eV$ and $-2.5~eV$ used in Fig.~\ref{figVOLTAGE}(a) and (b), respectively. 

One might expect that the conductances of two aforementioned magnetic configurations may strongly differ due to significant differences in their DOS, thereby leading to high magnetoresistance values defined as ($\sigma_{xx}^{2in2out}-\sigma_{xx}^{4in})/\sigma_{xx}^{4in}$), where $\sigma_{xx}^{4in}$ and $\sigma_{xx}^{2in2out}$ are the 
conductances for "4in" and "2in2out" magnetic configurations, respectively. We calculate the linear response conductance
using the Kubo formula~\cite{Kubo}
$$
\sigma_{xx}= \frac{\pi e^2 \hbar}{L} Tr 
\left[ \hat{v}_x \delta(E - \hat{H}) \hat{v}_x \delta(E - \hat{H}) \right],
$$
where $\hat{v}_x$ is the electron velocity operator along the applied voltage direction, $L \to \infty$ is the length of the system and 
$$
\delta(E - \hat{H})=-\frac{1}{\pi}\Im(\hat{G}^r(E))=\frac{i}{\pi}(\hat{G}^r(E)-\hat{G}^a(E))
$$

\begin{figure}
\includegraphics[width=7cm]{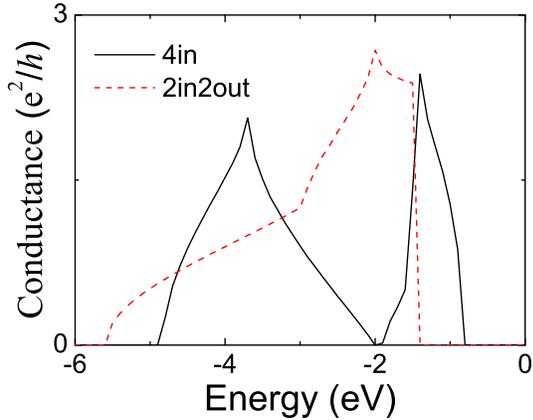}
\caption{\label{fig4} Conductance as a function of the Fermi level.}
\end{figure}

The calculated conductance of the system for "4in" and "2in2out" magnetic configurations are represented in Fig.~\ref{fig4}. As one can see, 
both $\sigma_{xx}^{4in}$ and $\sigma_{xx}^{2in2out}$ strongly depend on position of the Fermi energy $E_F$ and strongly correlate with DOS picture (see Fig.~\ref{figDOS}). 
For the two values of the Fermi energy used above, the MR ratio is found to be about 336~\% for the case when 
$E_F=-2.5~eV$ and becomes infinite for $E_F=-2~eV$.

One can note that similar behavior is expected to occur in systems with collinear antiferromagnetic structure: depending on the band structure and the band filling, the ferromagnetic state can be close in energy with the AF one, and an applied voltage could stabilize the ferromagnetic state. However in such a case, there is no spin torque since the structure is colinear, and no mechanism for spin reversal (except through spin waves). 
On the contrary, the mechanism proposed here neither require any defect, spin wave nor a fine tuning of any kind.
The noncollinear nature of the magnetic configurations considered provides a robust mechanism for microscopic magnetic switching.
Furthermore, this mechanism does not require spin polarized current injection.

In conclusion, we have shown that magnetoresistance of noncollinear frustrated bulk magnets may reach extremely 
high values and their magnetic configuration may be controlled by applied voltage. The proposed phenomenon 
is the bulk material analog of spin transfer torque used in layered spin-valve structures. The model chosen here used to 
prove the concept, in combination with inverse approach of designing the crystal structure from 
predefined electronic structure properties~\cite{Zunger}, opens the way towards experimental realization of the proposed 
phenomenon.
\begin{acknowledgments}
This work was supported by Chair of excellence Program of the Nanosciences Foundation in Grenoble.
\end{acknowledgments}

\begin{thebibliography}{2}

\bibitem{Baibich} 
M. Baibich et al, Phys. Rev. Lett. 61, 2472 (1988) 

\bibitem{Binash} 
G. Binasch, P.~Gr\"unberg, F.~Saurenbach and W.~Zinn, Phys. Rev. B 39, 4828 (1989)

\bibitem{Fert} 
A. Fert et al, Mat. Sci. Eng. B, 84, 1 (2001) 

\bibitem{Wolf}
S. Wolf, Science, 294, 1488 (2001)

\bibitem{Dieny}
B.~Dieny et al, Phys. Rev. B 43, 1297 (1991)

\bibitem{Slonczewski}
J. Slonczewski, JMMM 159, L1 (1996) 

\bibitem{Berger}
L. Berger, Phys. Rev. B 54, 9353 (1996)

\bibitem{Helmolt}
R. von Helmolt, J. Wecker, B. Holzapfel, L. Schultz, and K. Samwer, Phys. Rev. Lett. 71, 2331 (1993)

\bibitem{Jin}
S. Jin et al, Science, 264, 413 (1994)

\bibitem{gignoux1991}
D. Gignoux and D. Schmitt, J.Magn. Magn. Mater, \textbf{100}, 99 (1991) 
and J. Alloys and Comp. \textbf{326}, 143 (2001). 

\bibitem{auneau1995}
I. Auneau et al, Physica B \textbf{212}, 351 (1995)

\bibitem{schobinger2008}
P. Schobinger-Papamantellos et al, J. Phys. : condens Matter \textbf{20}, 195202 (2008)

\bibitem{burlet1981}
P. Burlet et al, Sol. St. Comm. \textbf{39}, 745 (1981)

\bibitem{gardner2009}
J. Gardner, M. J. P. Gingras, J. E. Greedan, arXiv:0906.3661; to appear in Rev. Mod. Phys

\bibitem{nakatsuji2006}
S. Nakatsuji et al, Phys. Rev. Lett. \textbf{96}, 087204 (2006)

\bibitem{Machida2007}
Y. Machida et al, Phys. Rev. Lett \textbf{98}, 057203 (2007)

\bibitem{kezsmarki2004}
I. K\'ezsm\'arki et al., Phys. Rev. Lett. {\bf 93}, 266401 (2004))

\bibitem{Lifshitz}
E. M. Lifshitz and L. P. Pitaevskii, \textit{Physical Kinetics, Course of Theoretical Physics}
(Pergamon, Oxford, 1981), Vol. 10

\bibitem{Meir}
Y. Meir and N.~S.~Wingreen, Phys. Rev. Lett. 68, 2512 (1992)

\bibitem{Caroli}
C. Caroli and R. Combescot and P. Nozieres and D. Saint-James, J. Phys. C: Solid St. Phys., \textbf{4}, 916 (1971)

\bibitem{Solovyev}
I.~V.Solovyev, Phys. Rev. B 67, 174406 (2003)

\bibitem{Kubo}
R. Kubo, J. Phys. Soc. Jpn, \textbf{12}, 570 (1957)

\bibitem{DOSdefin}
The Density of States is equal to $-1/\pi \sum_j \Im G_{jj}^R(E)$, where summation includes all atoms in the unit cell and spins.

\bibitem{Zunger}
A. Franceschetti and A. Zunger, Nature 402, 60 (1999)




\end{thebibliography}

\end{document}